\documentclass{PoS}

\newcommand{\ke}{K_{e2}}
\newcommand{\km}{K_{\mu 2}}
\def\NIMA#1#2#3{Nucl. Inst. Methods {\bf A#1} (#2) #3}

\title{Lepton universality test with $K^+\to l^+\nu$
decays at the NA62 experiment at CERN}

\ShortTitle{Lepton universality test with $K^+\to l^+\nu$ decays}

\author{\speaker{Evgueni Goudzovski}\\
        University of Birmingham\\
        E-mail: \email{eg@hep.ph.bham.ac.uk}}

\abstract{The NA62 experiment at CERN collected a large sample of
$K^+\to e^+\nu$ decays during a dedicated run in 2007, which allows
a precise test of lepton universality by measurement of the helicity
suppressed ratio $R_K = \Gamma(K^+\to
e^+\nu)/\Gamma(K^+\to\mu^+\nu)$. The preliminary result of the
analysis of a partial data sample of 51089 $K^+\to e^+\nu$
candidates is $R_K=(2.500\pm0.016)\times10^{-5}$, consistent with
the Standard Model expectation.}

\FullConference{2009 KAON International Conference KAON09,\\
         June 09 - 12 2009\\
         Tsukuba, Japan}

\begin{document}

\section*{Introduction}

The ratio of kaon leptonic decay rates $R_K=\Gamma(\ke^\pm)/\Gamma(\km^\pm)$
has been calculated with an excellent accuracy within the the Standard Model (SM)~\cite{ci07}:
$R_K^\mathrm{SM} = \left(m_e/m_\mu\right)^2
\left(\frac{m_K^2-m_e^2}{m_K^2-m_\mu^2}\right)^2 (1 + \delta
R_{\mathrm{QED}})= (2.477 \pm 0.001)\times 10^{-5}$,
where $\delta R_{\mathrm{QED}}=(-3.78\pm0.04)\%$ is a correction due to the inner bremsstrahlung (IB) $K_{\ell 2\gamma}$ process which is, unlike the structure dependent (SD) $K_{\ell 2\gamma}$ process, by definition included into $R_K$.
The factor $(m_e/m_\mu)^2$ accounts for the helicity
suppression of the $\ke$ decay due to the $V-A$ structure of the
charged weak current, and enhances sensitivity to non-SM effects.
In particular, enhancement of $R_K$ by a few
percent (relative) is possible in the MSSM with non-vanishing $e-\tau$ lepton mixing~\cite{ma06},
compatible to the presently known experimental constraints.

The current world average composed of three 1970s measurements~\cite{pdg} and a recent KLOE result~\cite{am09} is $R_K^{\rm WA}=(2.467\pm0.024)\times 10^{-5}$. It has 1\% relative precision and is compatible to the SM.
The NA62 experiment at CERN collected data in 2007--08 aiming at an
$R_K$ measurement with $0.4\%$ precision.
The preliminary result obtained with a partial data sample is presented here.

\section{Beams, detector and data taking}

The NA48/2 beam line and setup were used;
running conditions were optimized for the $\ke$ measurement in 2007
using the experience of earlier studies based on NA48/2 data sets~\cite{fa07}.

The beam line is capable of delivering simultaneous narrow momentum
band $K^+$ and $K^-$ beams; a central momentum 74~GeV/$c$ was
used in 2007. Momentum of the incoming kaon is not measured directly in every
event; the beam average monitored with
$K^\pm\to3\pi^\pm$ decays is used to reconstruct
$K_{l2}$ kinematics by missing mass $M_{\rm miss}$.
A narrow momentum spectrum ($\Delta p_K^{\mathrm{RMS}}/p_K\approx 2\%$)
is used to minimize the corresponding
contribution to the $M_{\rm miss}$ resolution.

The $K_{l2}$ decay signature consists of a single reconstructed
track. Since the incoming $K^+$ is not tracked, backgrounds induced
by the beam halo have to be considered. The performance of the muon
sweeping system results in lower background in $K_{e2}^+$ sample
($\sim 1\%$) than in $K_{e2}^-$ sample ($\sim 20\%$), therefore
$\sim 90\%$ of the data were taken with the $K^+$ beam only, and
small fractions were recorded with simultaneous beams and $K^-$ beam
only. The halo background is directly measurable using the samples
of reconstructed $K_{l2}$ candidates of the sign not present in the
beam.

Among the subdetectors located downstream a vacuum decay
volume, a magnetic spectrometer, a plastic scintillator hodoscope (HOD) and a liquid krypton electromagnetic calorimeter (LKr) are principal for the measurement. The spectrometer, used to detect charged products of kaon decays, is composed of four drift chambers (DCHs) and a dipole magnet.
The HOD, used to produce fast
trigger signals, consists of two planes
of strip-shaped counters. The LKr, used for
$\gamma$ detection and particle identification, is
an almost homogeneous ionization chamber, $27X_0$ deep, segmented
transversally into
13,248 cells (2$\times$2 cm$^2$ each),
and with no longitudinal segmentation.
A beam pipe traversing the centres of the detectors allows
undecayed beam particles and muons from decays of beam pions
to continue their path in vacuum.

A minimum bias trigger configuration is employed, resulting in high
efficiency with relatively low purity. The $\ke$ trigger condition
consists of coincidence of hits in the HOD planes (the so called
$Q_1$ signal) with 10 GeV LKr energy deposition. The $\km$ trigger
condition consists of the $Q_1$ signal alone downscaled by a factor
of 150. Loose upper limits on DCH activity are also applied.

Most data taking took place during four months in 2007. Two weeks of
data taking allocated in 2008 were used to collect special data
samples for studies of systematic effects. The present analysis is
based on $\sim 40\%$ of the 2007 data sample collected with the
$K^+$ beam only.

\section{Analysis strategy and event selection}

Monte Carlo (MC) simulations
are used to a limited extent only: 1) to evaluate a correction for
the difference of $\ke$ and $\km$ geometric
acceptances; 2) to simulate energetic
bremsstrahlung by a muon, which is not directly accessible experimentally as discussed below.

In order to compute geometrical acceptances, a
detailed Geant3-based MC simulation is employed. It includes
full detector geometry and material description, stray magnetic fields, local inefficiencies, misalignment,
detailed simulation of the beam line, and time variations
of the above throughout the running period. The $K_{\ell 2 (\gamma)}$ processes are simulated in one-photon approximation~\cite{ci07}. Unlike the KLOE analysis~\cite{am09}, the resummation of leading logarithms~\cite{ga06} is not
included into the simulation for the preliminary result.

The analysis strategy is based on counting the numbers of
reconstructed $\ke$ and $\km$ candidates collected simultaneously,
consequently the result does not rely on kaon flux measurement, and
the systematic effects due to the detector efficiency cancel to
first order. To take into account the momentum dependence of signal
acceptance and background level, the measurement is performed
independently in bins of reconstructed lepton momentum. The ratio
$R_K$ in each bin is computed as
\begin{equation}
R_K = \frac{1}{D}\cdot \frac{N(\ke)-N_{\rm B}(\ke)}{N(\km) - N_{\rm
B}(\km)}\cdot \frac{f_\mu\times A(\km)\times\epsilon(\km)}
{f_e\times A(\ke)\times\epsilon(\ke)}\cdot
\frac{1}{f_{\rm LKr}}, \label{RKexp}
\end{equation}
where $N(K_{\ell 2})$ are the numbers of selected $K_{\ell 2}$
candidates $(\ell=e,\mu)$, $N_{\rm B}(K_{\ell 2})$ are numbers of
background events, $f_\ell$ are efficiencies of $e$/$\mu$
identification criteria, $A(K_{\ell 2})$ are geometrical
acceptances computed with MC, $\epsilon(K_{\ell 2})$ are trigger
efficiencies, $f_{\rm LKr}$ is the global efficiency of the LKr
readout, and $D=150$ is the downscaling factor of the $\km$
trigger.

Due to topological similarity of $\ke$ and $\km$ decays,
a large part of the selection conditions is
common for both decays:
(1) exactly one reconstructed charged particle track;
(2) its momentum
$15~{\textrm{GeV}}/c<p<65~{\textrm{GeV}}/c$ (the lower limit is due
to the 10 GeV LKr energy deposit trigger requirement in $\ke$ trigger);
(3) extrapolated track impact points in DCH, LKr and HOD
are within their geometrical acceptances; (4) no LKr energy
deposition clusters with energy $E>2$~GeV and
not associated to the track to suppress background from other
kaon decays; (5) distance between the charged track and the nominal
kaon beam axis ${\rm CDA}<1.5$~cm, decay vertex longitudinal
position within the nominal decay volume
(the latter condition is optimized in each momentum bin).

The following two principal selection criteria are
different for the $\ke$ and $\km$ decays. $K_{\ell 2}$ kinematic identification
is based on the reconstructed squared missing mass assuming the track
to be an electron or a muon: $M_{\mathrm{miss}}^2(\ell) = (P_K - P_\ell)^2$,
where $P_K,P_\ell$ ($\ell = e,\mu$) are the four-momenta of the kaon
(average beam momentum assumed) and the lepton (electron or
muon mass assumed). A cut
$|M_{\mathrm{miss}}^2(e)|<M_0^2$ is applied to
select $\ke$ candidates,
and $|M_{\mathrm{miss}}^2(\mu)|<M_0^2$ for $\km$ ones, where
$M_0^2$ varies from 0.009 to 0.013~$(\mathrm{GeV}/c^2)^2$ among
track momentum bins depending on $M_{\mathrm{miss}}$ resolution.
Particle identification is based on the ratio $E/p$ of
track energy deposit in the LKr to its momentum
measured by the spectrometer.
Particles with $0.95<E/p<1.1$ ($E/p<0.85$) are identified as electrons (muons).

\section{Backgrounds}

\begin{figure}[tb]
\begin{center}
\resizebox{0.51\textwidth}{!}{\includegraphics{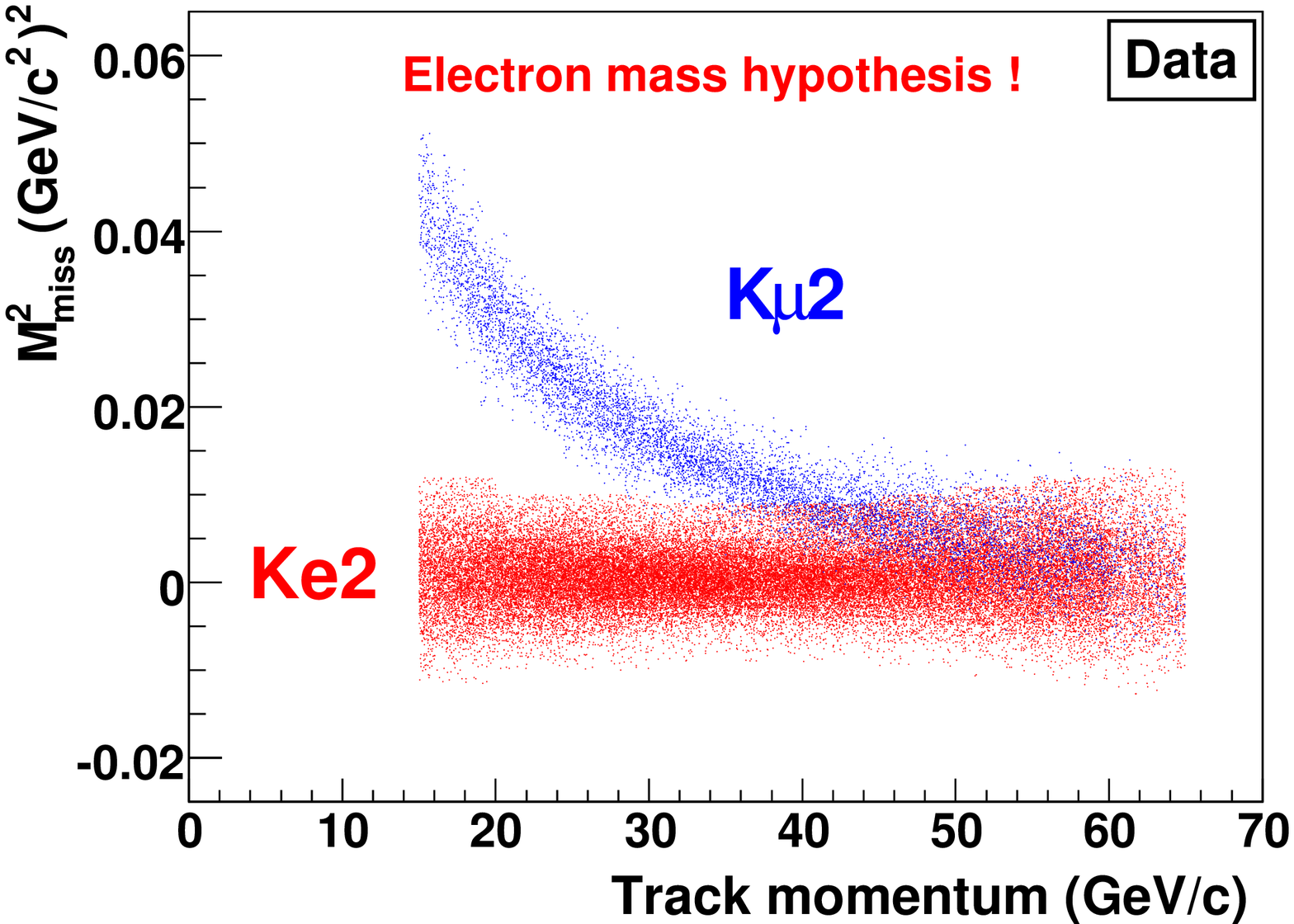}}%
\resizebox{0.49\textwidth}{!}{\includegraphics{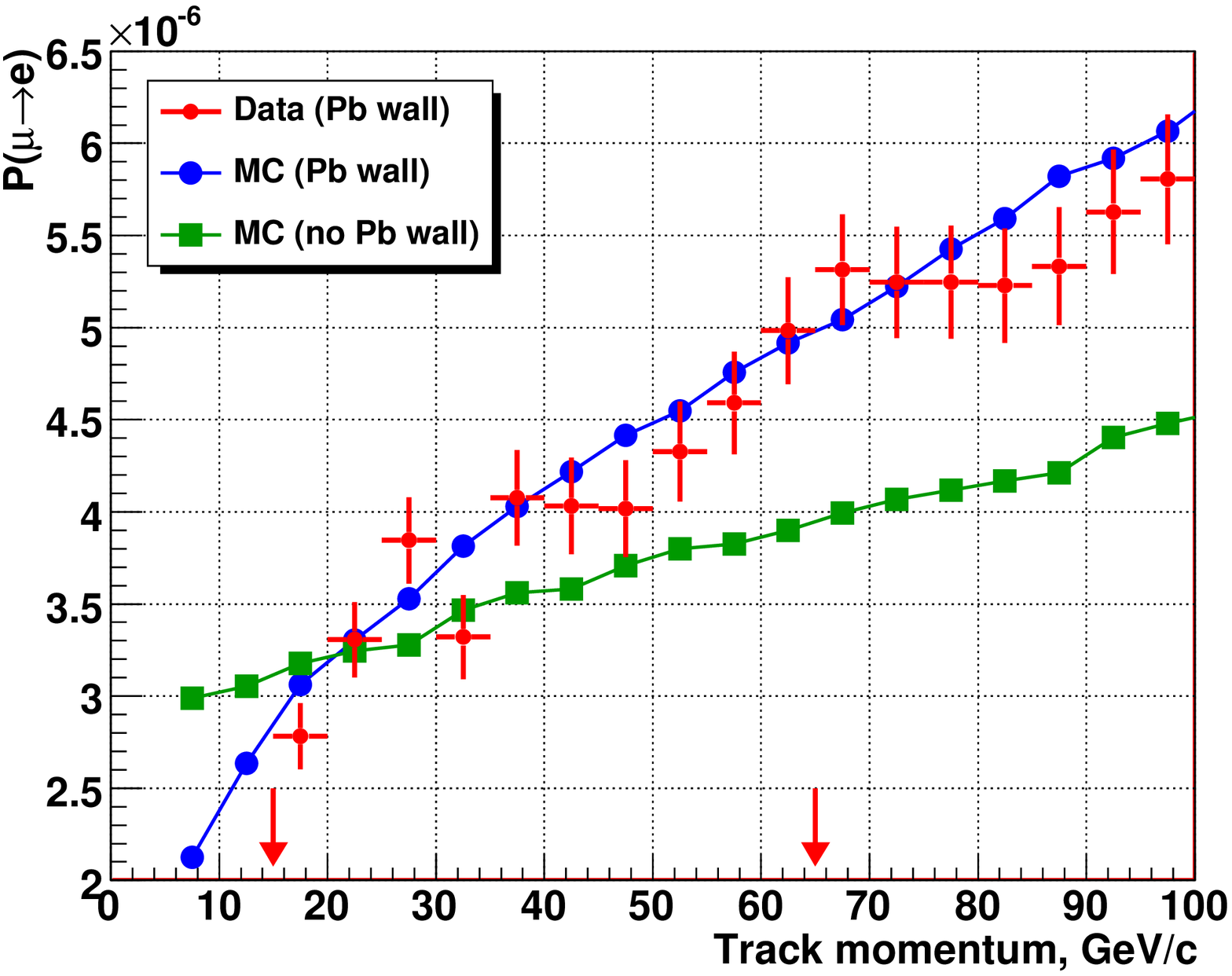}}
\put(-388,141){\large\bf (a)} \put(-60,144){\large\bf (b)}
\end{center}
\vspace{-7mm} \caption{(a) Missing mass squared in electron
hypothesis $M_{\rm miss}^2(e)$ vs track momentum for reconstructed
$K_{e2}$ and $K_{\mu2}$ decays: kinematic separation of $\ke$ and
$\km$ decays is possible at low track momentum only. (b) Measured
and simulated probability of muon identification as electron
$P(\mu\to e)$ vs track momentum: data with the Pb wall, MC
simulations with and without the Pb wall (signal region marked with
arrows).} \label{fig:pbwall}
\end{figure}

{\bf $K_{\mu2}$ decay} is the main background source
in the $\ke$ sample.
Kinematic separation of $\ke$ and $\km$ decays
is not achievable at high lepton momentum ($p>40$~GeV/$c$), as shown in Fig.~\ref{fig:pbwall}a.
The probability of muon identification as electron ($E/p>0.95$
due to `catastrophic' bremsstrahlung) is
$P(\mu\to e)\sim3\times10^{-6}$ in the NA62 experimental conditions,
non-negligible compared to $R_K^{\rm SM}=2.477\times 10^{-5}$.
Direct measurement of $P(\mu\to e)$ to $\sim 10^{-2}$ precision
is necessary for validation of a theoretical calculation of
the bremsstrahlung cross-section~\cite{ke97}
in the high $\gamma$ energy range, which is used to evaluate the $\km$ background.
Typical $\mu$ samples are affected by relatively large $(\sim10^{-4})$ electron contamination due to $\mu\to e$ decays in flight.
To collect pure $\mu$ samples,
a $\sim 10X_0$ thick lead (Pb) wall covering $\sim 20\%$
of the geometric acceptance was installed between
the HOD planes during a part of the data taking.
In the samples of tracks traversing the Pb and having $E/p>0.95$,
the electron component is suppressed to a level much below $P(\mu\to e)$
due to energy loss in Pb.

The momentum dependence of $P(\mu\to e)$ for muons traversing Pb was
measured with a data sample collected during a special 20h muon run,
and compared to the results of a dedicated Geant4-based MC
simulation involving standard muon energy loss processes and
bremsstrahlung according to~\cite{ke97}. The data/MC comparison
(Fig.~\ref{fig:pbwall}b) shows excellent agreement in a wide
momentum range within statistical errors, which validates the
cross-section calculation at the corresponding precision level. The
simulation shows that the Pb wall modifies $P(\mu\to e)$ via two
principal mechanisms: 1) muon energy loss in the Pb by ionization
decreasing $P(\mu\to e)$ and dominating at low momentum; 2)
bremsstrahlung in Pb increasing $P(\mu\to e)$ and dominating at high
momentum.

To estimate the $\km$ background contamination, the kinematic suppression
factor is computed with the standard setup simulation,
while the validated simulation of muon interaction
in the LKr is employed to account for $P(\mu\to e)$. The
uncertainty of the background estimate comes from the limited size of the data sample used to validate the simulation with the Pb wall.

{\bf $K_{\mu2}$ decay followed by $\mu\to e$ decay}: energetic forward electrons contributing to background are suppressed according to the Michel distribution as muons from $\km$ decays are fully polarised.

{\bf $K_{e2\gamma}$ (SD) decay}, a background by $R_K$ definition,
has a rate similar to that of $\ke$: experimentally
${\rm BR}=(1.52\pm0.23)\times10^{-5}$~\cite{pdg}. Theoretical BR
calculations have similar precision,
depending on the form
factor kinematic dependence model. Energetic electrons
($E_e^*\gtrsim 230$~MeV in $K$ frame) with $\gamma$ escaping detector acceptance
contribute to the background.
MC background estimation has a 15\% uncertainty due to
limited knowledge~\cite{pdg} of the process. A recent precise measurement
by KLOE~\cite{am09}, published after announcement of the NA62 preliminary result, is not taken into account.

{\bf Beam halo} background in the $\ke$ sample
induced by halo muons undergoing $\mu\to e$ decays in flight
is measured directly using the $K^-$ data samples.
Background rate and kinematical distribution are
qualitatively reproduced by a muon halo simulation.
The uncertainty is due to the limited size of the $K^-$ sample.
Beam halo is the only significant background source in the $\km$ sample, measured
to be $0.25\%$ (with a negligible uncertainty) with the same technique as for $\ke$ decays.

\begin{figure}[tb]
\begin{center}
\resizebox*{0.47\textwidth}{!}{\includegraphics{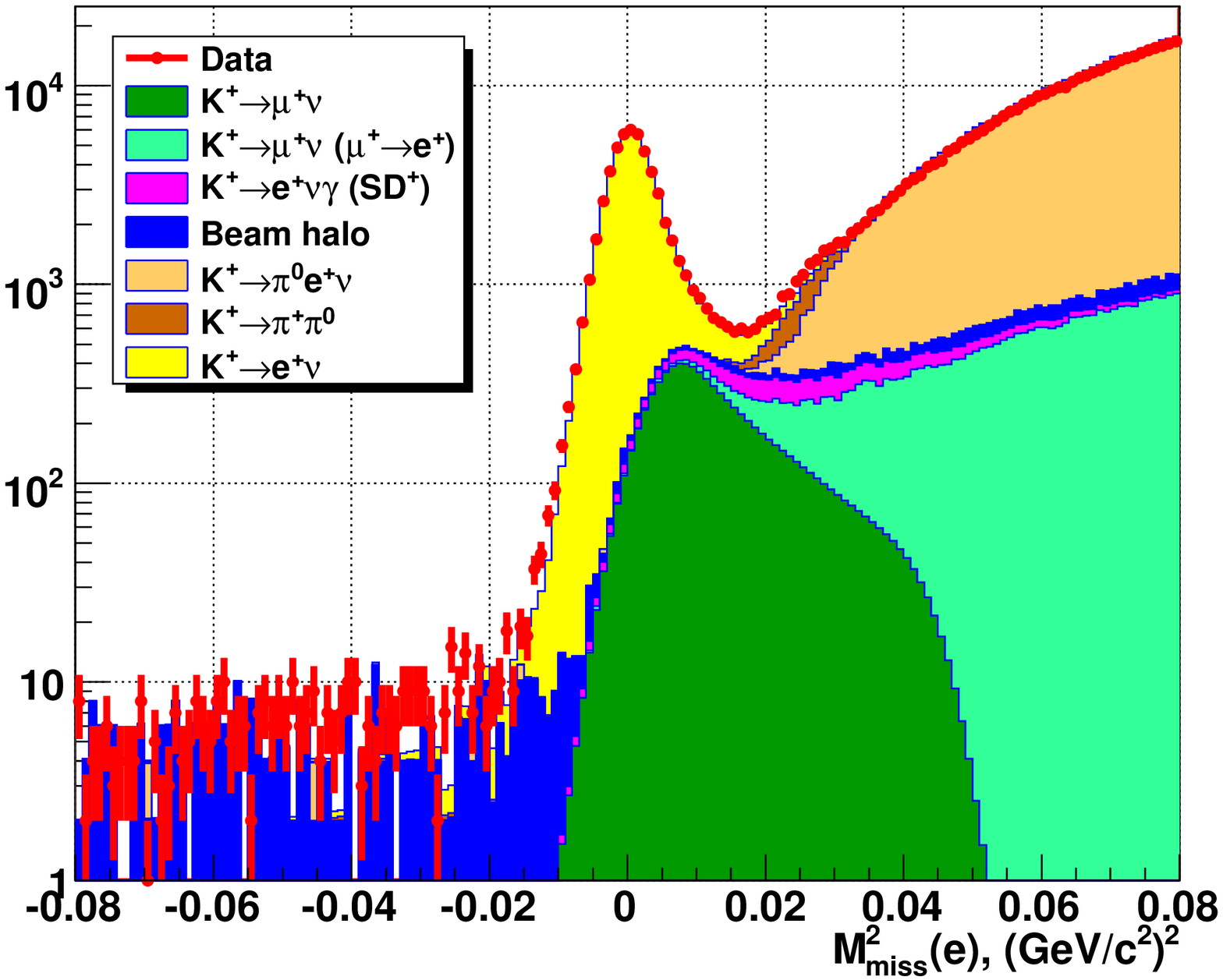}}~~
\resizebox*{0.45\textwidth}{!}{\includegraphics{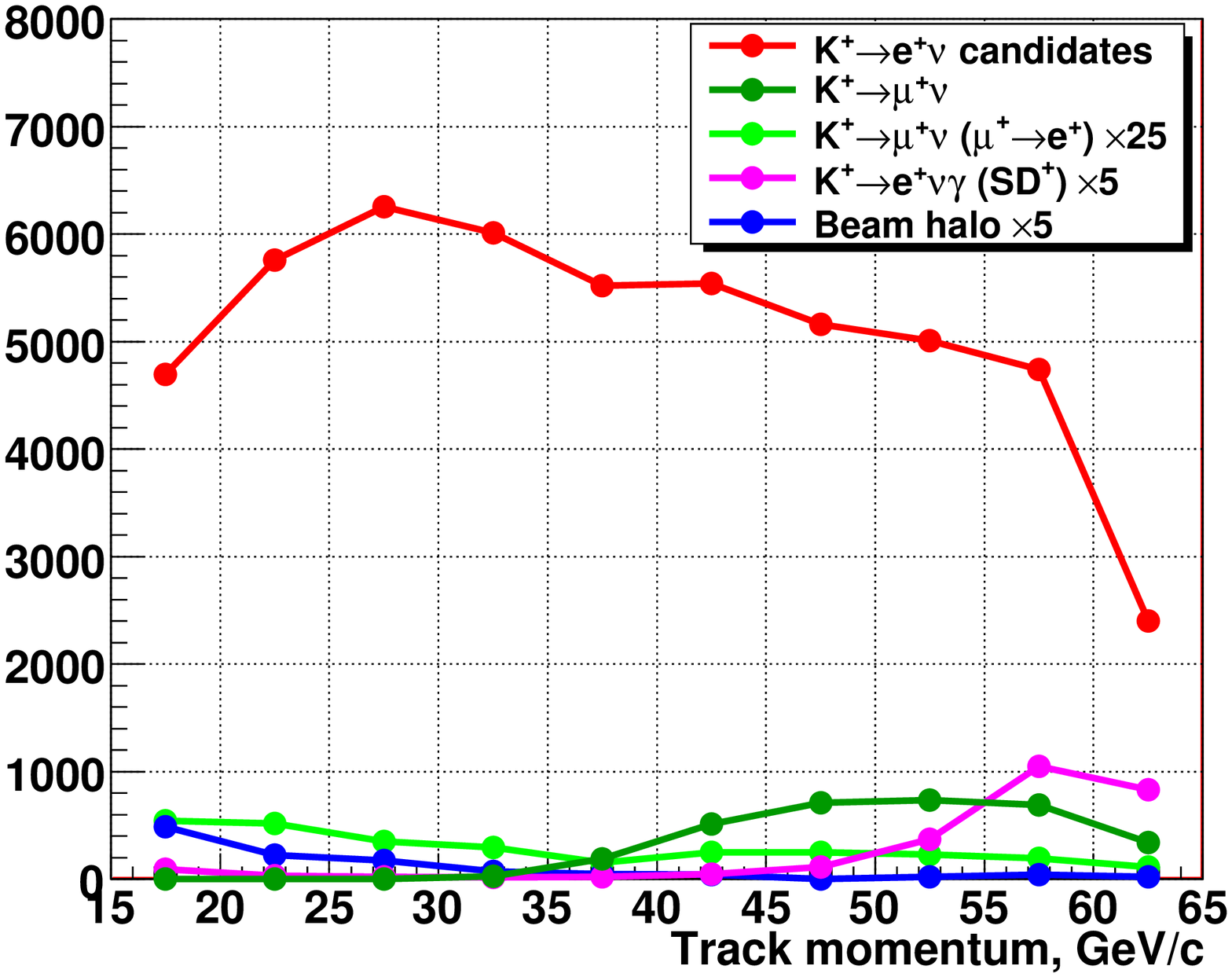}}
\end{center}
\vspace{-7mm}
\caption{(a) Reconstructed squared missing mass distribution
$M_{\mathrm{miss}}^2(e)$ for the $\ke$ candidates: data (dots) presented as sum of signal and backgrounds (filled areas). (b) Numbers of $\ke$ candidates and background events in track momentum bin.}
\label{fig:mmiss}
\end{figure}

\begin{table}
\begin{center}
\begin{tabular}{lc|lc|lc}
\hline Source & $N_B/N_{tot}$ & Source & $N_B/N_{tot}$ & Source & $N_B/N_{tot}$\\
\hline
$\km$               & $(6.28\pm0.17)\%$ & $K_{e2\gamma}$~(SD) & $(1.02\pm0.15)\%$ & $K_{e3}$   & $0.03\%$\\
$\km~(\mu\to e)$    & $(0.23\pm0.01)\%$ & Beam halo  & $(1.45\pm0.04)\%$          & $K_{2\pi}$ & $0.03\%$\\
\hline
\multicolumn{6}{c}{Total background: $(8.03\pm0.23)\%$}\\
\hline
\end{tabular}
\end{center}
\vspace{-6mm} \caption{Summary of the background sources in the $\ke$ sample.}
\label{tab:bkg}
\end{table}

The number of $K_{\ell 2}$ candidates is $N(\ke)=51,089$ (about four
times the statistics reported by KLOE~\cite{am09}) and
$N(\km)=15.56\times 10^6$. The $M_{\mathrm{miss}}^2(e)$
distributions of data events and backgrounds are presented in
Fig.~\ref{fig:mmiss}a. Backgrounds integrated over track momentum
are summarized in Table~\ref{tab:bkg}; their distributions over
track momentum are presented in Fig.~\ref{fig:mmiss}b.\vspace{-2mm}

\section{Systematic uncertainties and results}

{\bf Electron identification efficiency} is measured directly as a
function of track momentum and LKr impact point using pure samples
of electrons obtained by kinematic selection of $K^+\to\pi^0 e^+\nu$
decays collected concurrently with the $\ke$ sample, and
$K^0_L\to\pi^\pm e^\mp\nu$ decays from a special 15h $K^0_L$ run.
The $K^+$ and $K^0_L$ measurements are in good agreement. The
measured $f_e$ averaged over the $\ke$ sample is 99.2\%, with a
precision better than 0.05\%. Muon identification inefficiency is
negligible.

{\bf The geometric acceptance correction} $A(\km)/A(\ke)$ depends on
the radiative $K_{e2\gamma}$ (IB) decays. A conservative systematic
uncertainty is attributed to approximations used in the
$K_{e2\gamma}$ IB simulation, which follows~\cite{ci07}. The
resummation of leading logarithms~\cite{ga06} is not taken into
account, however no systematic error is ascribed due to that. An
additional systematic uncertainty reflects the precision of beam
line and apparatus description in the MC simulation.

{\bf Trigger efficiency} correction
$\epsilon(\ke)/\epsilon(\km)\approx 99.9\%$ accounts for
the fact that $\ke$ and $\km$ decay modes are
collected with different trigger conditions: the $E>10$~GeV LKr energy
deposition signal enters the $\ke$ trigger only. A conservative systematic
uncertainty of 0.3\% is ascribed due to effects of trigger dead time which affect the two modes differencly. {\bf LKr global readout efficiency} $f_{\rm LKr}$
is measured directly to be $(99.80\pm0.01)\%$ using an independent LKr readout.

\begin{figure}[tb]
\begin{center}
{\resizebox*{0.43\textwidth}{!}{\includegraphics{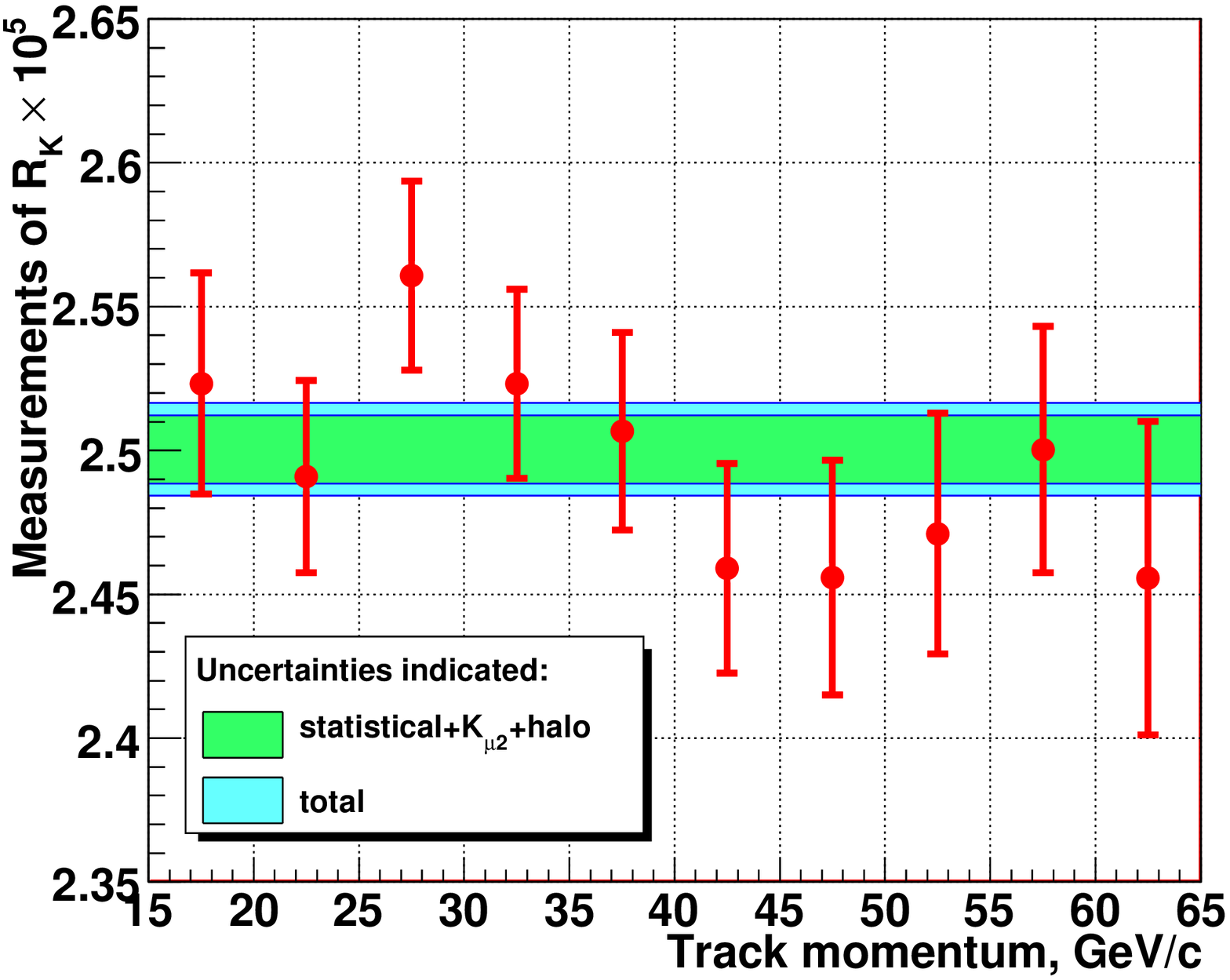}}}~~
{\resizebox*{0.43\textwidth}{!}{\includegraphics{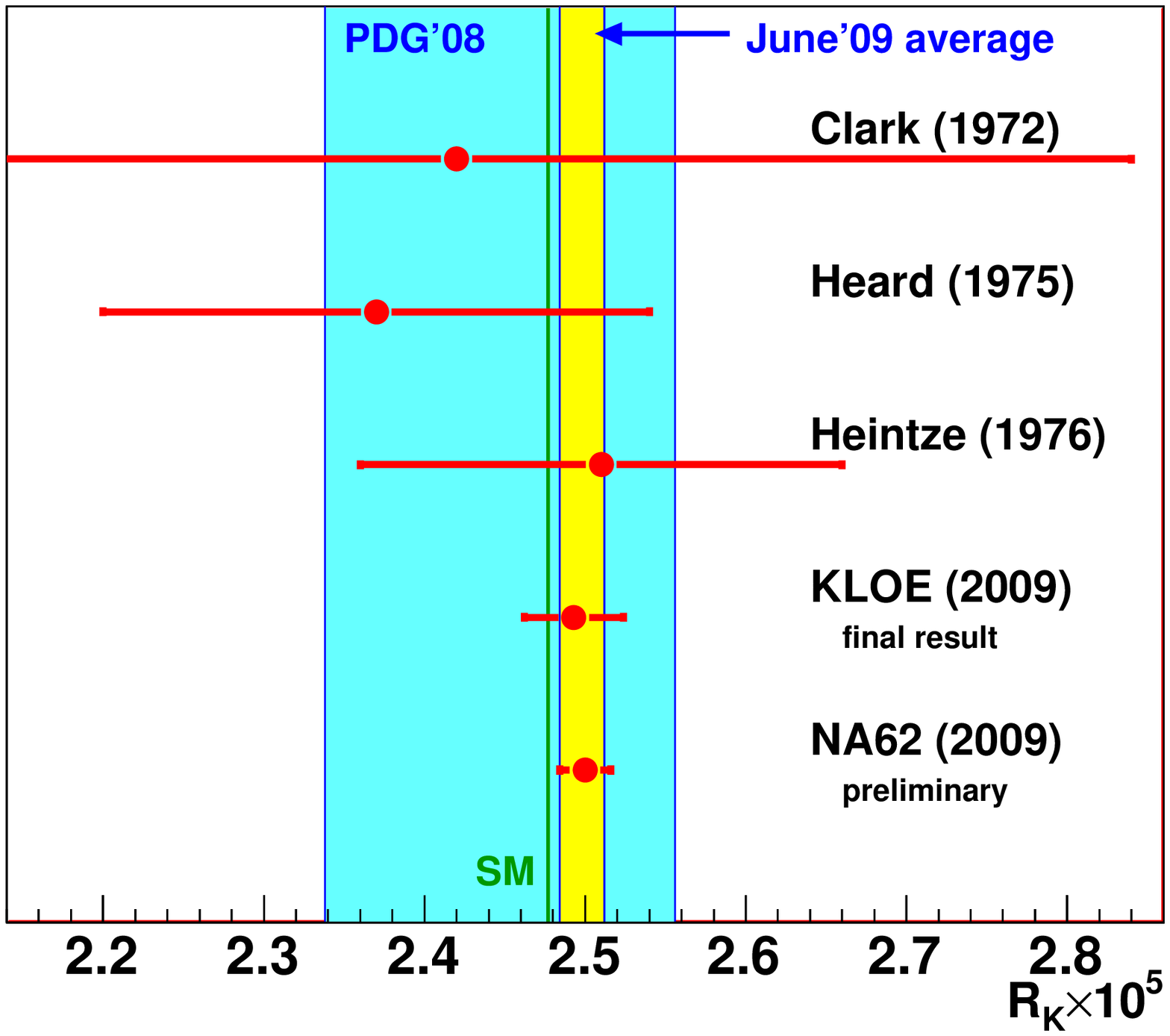}}}
\end{center}
\vspace{-9.5mm} \caption{(a) $R_K$ in track momentum bins.
(b) Summary of $R_K$ measurements.}
\label{fig:rk}
\end{figure}

The independent measurements of $R_K$ in track momentum bins with
are presented in Fig.~\ref{fig:rk}a. The uncertainties are
summarised in Table~\ref{tab:err}. The preliminary NA62 result is
$R_K=(2.500\pm0.012_{\rm stat.}\pm0.011_{\rm syst.})\times 10^{-5} =
(2.500\pm0.016)\times 10^{-5}$, consistent with the SM expectation.
The whole 2007--08 data sample will allow pushing the uncertainty of
$R_K$ down to 0.4\%. A summary of $R_K$ measurements is presented in
Fig.~\ref{fig:rk}b: the new world average is $(2.498\pm0.014)\times
10^{-5}$.\vspace{-2mm}

\begin{table}
\begin{center}
\begin{tabular}{lc|lc|lc}
\hline Source & $\delta R_K\times 10^5$ & Source & $\delta R_K\times 10^5$ & Source & $\delta R_K\times 10^5$\\
\hline
Statistical         & 0.012 & Beam halo      & 0.001 & Geom. acceptance  & 0.002\\
$\km$               & 0.004 & Electron ID    & 0.001 & Trigger dead time & 0.007\\
$K_{e2\gamma}$ (SD) & 0.004 & IB simulation  & 0.007\\
\hline
\end{tabular}
\end{center}
\vspace{-6mm} \caption{Summary of uncertainties of $R_K$:
statistical and systematic contributions.}
\label{tab:err}
\end{table}

\end{document}